\documentstyle[twoside,fleqn,espcrc2]{article}

\newcommand{\lan}{{\langle}}
\newcommand{\ran}{{\rangle}}
\newcommand{\be}{\begin{equation}}
\newcommand{\ee}{\end{equation}}
\newcommand{\bea}{\begin{eqnarray}}
\newcommand{\eea}{\end{eqnarray}}
\newcommand{\da}{\dagger} 

\newcommand{\al}{\alpha}
\newcommand{\xm}{x+\hat \mu}
\newcommand{\xn}{x+\hat \nu}

\newcommand{\fhi}{\varphi}

\newcommand{\xmm}{x-\hat \mu}

\newcommand{\ra}{\rightarrow}

\newcommand{\AmS}{{\protect\the\textfont2
  A\kern-.1667em\lower.5ex\hbox{M}\kern-.125emS}}

\title{Correlation Functions from Quantum Cluster Algorithms: 
    Application to U(1) Quantum Spin and Quantum Link Models
\thanks{Work supported in part by funds provided by the 
U.S. Department of Energy (D.O.E.) under cooperative research agreement 
DE-FC02-94ER40818.}}

\author{A. Tsapalis \address{Center for Theoretical Physics, 
Laboratory for Nuclear Science and 
Department of Physics \\ 
Massachusetts Institute of Technology (MIT), Cambridge, MA 02139}}

\begin{document}

\begin{abstract}
We demonstrate how correlation functions for non-diagonal operators can be
measured with the loop-cluster algorithm for quantum spin models. We 
introduce the $U(1)$ quantum link model and present the construction of
a cluster algorithm for the model. We further show how an improved estimator
for Wilson loops can be realized with the flux-cluster algorithm.

\end{abstract}

\maketitle

\section{INTRODUCTION}

Cluster algorithms have been established as very efficient tools for the
numerical simulation of both classical and quantum spin systems.
In particular, for quantum spin systems, the loop-cluster algorithm  
\cite{Eve93} has allowed very efficient studies of the 2-d 
Heisenberg antiferromagnet
utilizing a continuous Euclidean time formulation \cite{Bea96}.
An interesting development is the
observation \cite{Bro98} that non-diagonal
correlation functions can be measured  with the loop-cluster algorithm.
The methodology is carried over to the numerical study of the $U(1)$
quantum link model (QLM) which is a new, discrete formulation of Wilson's gauge
theory \cite{Cha97}. A flux-cluster algorithm has been constructed 
and an improved estimator for Wilson loops is shown to exist.

\section{THE LOOP-CLUSTER ALGORITHM} 

Consider the 2-d quantum $XY$ model, defined by
the Hamiltonian
\be
\hspace{1.0cm} H = -J \sum_{x,\mu=1,2} [S^1_x S^1_{\xm} + S^2_x S^2_{\xm}] \,,
\ee
which couples nearest-neighbor quantum spin operators $\vec{S}_x$. 
The quantum partition function 
\be
\label{ZS}
\hspace{2.0cm} Z = \mbox{Tr}\exp(-\beta H)
\ee
is formulated as a (2+1)-d path integral with the system evolving in 
Euclidean time with extent $\beta$. Time is discretized in small steps of size 
$\epsilon$ and the Hamiltonian is decomposed in an even-odd
pattern \cite{Bea96} such that the transfer matrix between two
subsequent time slices is a product of mutually commuting elementary
transfer matrices
\be
\hspace{1.0cm} {\cal T}=\exp(\epsilon J[S^1_x S^1_{\xm} + S^2_x S^2_{\xm}])\,.
\ee
A complete set of states is inserted in each time-slice to form the path
integral. 
For $s$=1/2 spins, the $4\times 4$ matrix elements of ${\cal T}$ between 
the eigenstates --- for example in the $S^3$ basis ---
of the two spins, define the Boltzmann weights
for the interaction between two pairs of 
spin states in adjacent time slices. The loop-cluster algorithm
starts on a random spin state and examines the weight of the interacting 
quartette of spins. It joins probabilistically the spin to one of the
other spins and the process continues until a loop of spin states is formed.  
The loop is the worldline of evolution of a spin state in the (2+1)-d
periodic volume. The cluster rules are ergodic and obey detailed balance.
The spin states in the cluster are then flipped resulting in a very
efficient update of the system. It is shown in \cite{Bro98} that the quantum 
partition function is mapped to a quantum random cluster model similarly
to the Fortuin-Kasteleyn mapping for the Potts model.
The mapping holds at the operator level and therefore the loop-cluster 
has geometrical properties independent of the chosen basis. 
Due to the discreteness of the Hilbert space, 
the continuum limit $\epsilon \ra 0$ can be taken and the algorithm can be
implemented in the continuum of the Euclidean time direction \cite{Bea96}.

The expectation value of a Green's function like
\be
\label{greens}
\hspace{1.0cm} \lan S^1_x S^1_y \ran = 
\frac{1}{Z} \mbox{Tr}[S^1_x S^1_y \exp(-\beta H)]\;,  
\ee
appears impossible to measure
from the ensemble of configurations that contribute to
$Z$. This is because the insertion of the non-diagonal operator
$S^1$ creates a configuration that never contributes to the 
partition function. 
The important observation made in \cite{Bro98} 
is that the loop-cluster algorithm generates configurations
that can be thought as contributing to (\ref{greens}) also. Indeed, consider
the loop which passes from the sites $x$ and $y$ on a fixed time-slice.
We can imagine that we cut the loop on this time-slice and flip
only half of the cluster. Since $S^1$ flips the $S^3$ eigenstates, 
this process precisely contributes to (\ref{greens}). We therefore
generate the loops with the algorithm and then perform cuts on all time-slices
which dissect the loop in two pieces. If each half can be flipped 
independently,
a +1 contribution is registered for the correlation function between the
cutting sites. In this way, an {\it improved estimator} for the non-diagonal
Greens's function (\ref{greens})
has been realized which does not suffer from any sign cancelations.

\section{THE U(1) QUANTUM LINK MODEL}

A new formulation for lattice gauge theories has been proposed in \cite{Cha97}.
Here we study the pure $U(1)$ gauge theory on the lattice. We 
consider the quantization of the links, i.e. the promotion of 
the classical link field
$u_{x,\mu}=\exp(i\fhi_{x,\mu})$ to a quantum link operator $U_{x,\mu}$ 
acting on the link-based Hilbert space. The $U(1)$ QLM is 
defined through the 4-d Hamiltonian
\be
\;\;\;\;\;\; H = -J \sum_{x,\mu \not= \nu}
[U_{x,\mu}U_{\xm,\nu}U^{\da}_{\xn,\mu}U^{\da}_{x,\nu}]
\ee
which is invariant under the local $U(1)$ transformations at the left and right
ends of the link
\bea
U'_{x,\mu}&=& \prod_y \exp(i\al_y G_y)U_{x,\mu} \prod_z \exp(-i\al_z G_z) 
\nonumber  \\
&=& \exp(i\al_x)U_{x,\mu} \exp(-i\al_{\xm} )\;.
\eea
The transformation law results in the 
following commutation relations between the
local generator of the symmetry $G_x$ and the link operators 
\be
\hspace{1.2cm} [G_y,\,U_{x,\mu}]=(\delta_{y,x}-\delta_{y,\xm})U_{x,\mu} \;.
\ee
These relations are satisfied if we represent the link operators and the 
$U(1)$ generator as
\bea
\hspace{1.4cm} U_{x,\mu} = S^1_{x,\mu} + i S^2_{x,\mu} 
= S^+_{x,\mu} \\ \nonumber
G_x = \sum_{\mu}( S^3_{x,\mu} - S^3_{\xmm,\mu} )\;,
\eea
where $S^a_{x,\mu}$ is a spin operator associated with a link, with the
usual commutation relations
\be
\hspace{1.3cm} [S^a_{x,\mu},\,S^b_{y,\nu}] = i \delta_{xy}\delta_{\mu \nu}
\epsilon^{abc} S^c_{x,\mu}\,.
\ee
The operator $S^3_{x,\mu}$ represents the electric flux on the link and 
$U_{x,\mu}$ acts as a flux raising operator. The Hilbert space per link is
the space of the local $SU(2)$ algebra representation
and it is therefore finite.
The quantum partition function 
\be
\label{Z}
\hspace{2.2cm} Z = \mbox{Tr}\exp(-\beta H)
\ee
is formulated as a (4+1)-d path integral with the system evolving in a 
fifth dimension of extent $\beta$. 
When $\beta$ exceeds a critical 
value, the excitations of the model are in the Coulomb phase of the
5-d Abelian gauge theory. Due to the infinite correlation length in the
5-d Coulomb phase, if the extent of the fifth dimension $\beta$ is finite, 
the theory will dimensionally reduce to the 4-d Abelian gauge theory.
The phase transition of Wilson's $U(1)$ gauge theory can be studied from the
growth of the correlation length near the critical fifth dimension extent of 
the $U(1)$ QLM. Even spin-1/2 quantum links may suffice for 
the dimensional reduction --- more on the phase diagram of the model for
general spin in \cite{Cha98}.

\section{THE FLUX-CLUSTER ALGORITHM} 

A cluster algorithm has been constructed for the spin-1/2 $U(1)$ QLM
\cite{lat97,Bea98}. The fifth dimension in (\ref{Z})
is discretized in small steps of size $\epsilon$
and a complete set of states is inserted at
each time-slice. Each plane of the 4-d lattice Hamiltonian is decomposed
in a checker board fashion of even and odd plaquettes. Only one type of
plaquettes is allowed to interact between two subsequent time-slices so that
the full transfer matrix is a product of commuting 
transfer matrices
\bea
\hspace{0.7cm} {\cal T} &=& \exp(\epsilon J [U_{x,\mu}U_{\xm,\nu}
U^{\da}_{\xn,\mu} U^{\da}_{x,\nu} \\ \nonumber
&+& U_{x,\nu}U_{\xn,\mu}U^{\da}_{\xm,\nu}U^{\da}_{x,\mu}])\,.
\eea
The $16\times 16$ matrix elements of ${\cal T}$ define the
Boltzmann weights of the checker boarded cubes carrying the interaction of 
8 electric flux states $e_{x,\mu,t}$ in the path integral
\be
\hspace{1.1cm} Z = \prod_{x,\mu,t} \sum_{e_{x,\mu,t}=\pm1/2} \exp(-S[e])\;.
\ee
The flux-cluster algorithm builds a cluster of links in the 5-d volume and
updates the system by flipping the electric flux on the links which belong  
to the cluster. The cluster starts on a random link and probabilistically
assigns some of the 8 links of the interaction cube in the cluster. 
The cluster rules are ergodic and obey detailed balance.
The process continues through the 5-d volume until the cluster is completed.
The cluster is a 2-d surface embedded in the 5-d volume and is the 
worldsheet of the evolution in the fifth dimension of open
and closed electric flux strings. Open flux strings are allowed because the
Gauss law is not enforced in (\ref{Z}). Due to the discreteness of the flux,
the algorithm can operate directly in the continuum limit 
$\epsilon \ra 0$ of the Euclidean time direction \cite{Bea96,Bea98}.

The order parameter for the gauge theory is the expectation value 
of Wilson loops $W_{{\cal C}} $ which is the product of quantum link
operators $U_{x,\mu}$ along some closed curve ${\cal C}$ in the 4-d
lattice at a fixed value of $t$. The link operators on the curve raise the
electric flux and therefore the configurations which are generated by the
algorithm will never contribute to the expectation
value
\be
\hspace{1.2cm} \lan W_{{\cal C}} \ran =\frac{1}{Z}
\mbox{Tr}[W_{{\cal C}} \exp(-\beta H)]\,.
\ee
The flux-cluster algorithm provides the solution again. Each cluster which is
generated and flipped contributes to $Z$, but it may also contribute to
$\lan W_{{\cal C}} \ran$. The operator
$W_{{\cal C}}$ flips the orientation of the electric flux along the curve
${\cal C}$. If the cluster is cut along ${\cal C}$ and the flux is flipped 
on only one of the pieces, a +1 contribution to $\lan W_{{\cal C}}\ran $ is 
obtained. Notice that if the cluster is connected to itself by 
wrapping over the periodic fifth dimension, i.e. has the topology of a torus,
it cannot be cut in two disjoint pieces along a circle and
does not contribute to $\lan W_{{\cal C}}\ran $. With the flux-cluster
algorithm therefore, a lot of information can be collected by examining the
topology of the cluster and cutting the cluster along fixed time-slices. 
Each closed curve cut which allows independent flips of the two parts
offers a +1 contribution to the corresponding Wilson loop. We have
therefore realized an {\it improved estimator} with no sign cancelations 
for the order parameter of the Abelian gauge theory.

In conclusion, the quantum link formulation and the cluster algorithm
have provided a complete setup for an efficient study of the Abelian 
lattice gauge theory.



\section*{ACKNOWLEDGMENTS}
I wish to thank B. B. Beard, R. Brower, S. Chandrasekharan and U.-J. Wiese 
for the very enjoyable collaboration on this research.

\end{document}